# Single-beam driven rotational manipulation for high-resolution 3D cellular morphology reconstruction

**Yi-Wei Pan[1], Yi-Jing Wu[1], Zi-Qiang Wang[1], Yin-Mei Li,[1,2] and Lei Gong[1, 2,★]**

*[1]Department of Optics and Optical Engineering, University of Science and Technology of China, Hefei 230026, China*

*[2]Hefei National Laboratory for Physical Sciences at the Microscale, University of Science and Technology of China, Hefei 230026, China*

*★E-mail: leigong@ustc.edu.cn*

**The acquisition of multi-view information of cells is essential for accurate 3D reconstruction of their structures. Rotational manipulation of cells has emerged as an effective technique for obtaining such data. However, most reported methods require a trade-off between manipulation flexibility and system complexity These limitations significantly hinder their practical applicability. Recently, a novel approach has been proposed that enables simultaneous trapping and arbitrary-angle rotation of cells using a single optical beam carrying spin angular momentum (SAM). This method offers improved stability and manipulation flexibility, a simplified experimental setup, and supports coaxial alignment of the imaging and optical paths. In this paper, we employed this method to rotate cells and acquire multi-view images. Furthermore, we present a complete 3D reconstruction workflow, and validate the performance of the proposed method through the reconstruction of Punica granatum pollen cells and Prunus cerasifera cells. Our methods pave the way for 3D**

**reconstruction of microscopic biological specimens, including but not limited to cells.**

## INTRODUCTION

Reconstructing 3D model of cells is crucial for studying its morphological and structural characteristics, which play an important role in biological and medical research. Accurate reconstruction requires the acquisition of multi-view images of individual cell. Moreover, to avoid a decrease in axial resolution due to missing cones[1], it is essential to obtain a sufficient number of views from different angles.

The most straightforward approach to obtaining multi-view images of cells is to load a cell into a microcapillary[2, 3] or a rotating tip[4, 5]. However, such techniques typically require customized sample stages, which reduce experimental throughput and complicate sample mounting and fixation, thereby limiting their applicability to live-cell imaging. Recently, several methods based on acoustic field[6, 7], magnetic field[8], optofluidic[9] and optical field[10-12] have been reported. However, each of these approaches presents its own limitations in practical applications, and most require relatively complex setups. For example, acoustic field-based methods are more suitable for manipulating larger (millimeter-scale) particles, but their stability decreases significantly when applied to smaller particles such as cells. Magnetic field-based methods generate relatively weak forces and are primarily used in single-molecule manipulation. When applied to cells, they often require pre-treatment of the sample or the introduction of magnetic particles into the medium. Compared to these methods, optical field-based approaches are better suited for cell manipulation, and several methods for rotating cells have already been proposed. For example, employing a dual beam fiber optical trap to manipulate and by rotating structured modes to achieve cell rotation. However, such methods require the assistance of microfluidic and can only induce rotation along the optical axis. This not only increases the complexity of the experimental setup but also restricts the camera to observing the particle rotation only in the transverse direction, perpendicular to the optical axis. Recently, tomographic moulds[11] have been reported, which enable rotation along arbitrary directions by dynamically shaping the trapping beam. However, this method relies on rotating the trapping field, masking the rotation behavior closely dependent on the 3D refractive index distribution of the cell. In addition, some methods based on spin angular momentum (SAM), such as T-spin[13, 14], have also been reported. But these can only rotate the particles along a fixed direction, thereby limiting the observations direction of the camera. In addition, some methods that use the instability of optical tweezers or external flow fields to induce rotation have also been proposed. However, these suffer from low stability and limited control, making them less suitable for precise 3D imaging applications.

Recently, a novel method capable of simultaneously trapping and rotating cells along arbitrary directions along arbitrary directions using a single beam has been reported[15]. This method controls the SAM of a tightly focused light field by modulating the incident light. By transferring SAM to the cell, light can exert a torque, thereby enabling rotational manipulation. In contrast to conventional methods, this method enables cell rotation around arbitrary axes by controlling the 3D SAM vector of the focused field. Moreover, by adjusting the SAM to induce rotation around a transverse axis, the camera can observe the particle along a direction parallel to the optical axis, aligning with conventional microscopic imaging systems. In this paper, we exploited this method for in situ 3D imaging of freestanding microparticles. A neural network-based approach was employed to process a series of multi-view images captured by the camera, obtaining corresponding masks for each image. The 3D model of individual cells was reconstructed from multiple angles using the visual hull algorithm[16]. Moreover, we experimentally validated the proposed workflow by reconstructing the 3D models of Punica granatum pollen cells and Prunus cerasifera cells.

**RESULTS**

**Concept and principle**

Fig .1 illustrates the principle of our method. The SAM is a fundamental dynamical property of the light field that reflects the specific vector nature of electromagnetic field. When a light field interacts with matter, its SAM can be transferred to the object, inducing it rotation. In classical electrodynamics, we can calculate the SAM density for a monochromatic field by:

$$\vec{\mathbf{S}} = \frac{1}{2\omega_0} \text{Im}[\varepsilon_0 \overrightarrow{\mathbf{E}^*} \times \vec{\mathbf{E}}].$$ (1)

Where $\omega_0$ is the angular frequency of the light, $\varepsilon_0$ is the vacuum permittivity, $\text{Im}(\bullet)$ denotes the operator that extracts the imaginary part of a complex number, and the $\overrightarrow{\mathbf{E}^*}$ is the complex conjugate of the electric field $\vec{\mathbf{E}}$. From this equation, it is evident that the SAM density for a linearly polarized monochromatic plane wave is zero. However, this is not the case for circularly polarized light. A circularly polarized plane wave can be expressed as:

$$\vec{\mathbf{E}} = \frac{\vec{e}_x + i\sigma \vec{e}_y}{\sqrt{2}} \mathbf{E}_0 \, e^{i\vec{k}\cdot\vec{x} - i\omega_0 t}.$$ (2)

Where $\vec{\mathbf{k}}$ is wavevector that determines the propagation direction of plane wave, $i$ is the imaginary unit, $\vec{\mathbf{e}}_x$ and $\vec{\mathbf{e}}_y$ are unit vectors lying in the plane perpendicular to wavevector $\vec{\mathbf{k}}$, satisfying $\vec{\mathbf{e}}_x \times \vec{\mathbf{e}}_y \parallel \vec{\mathbf{k}}$. The scalar $\sigma$ takes the value $\pm 1$ and $E_0$ is the amplitude of plane wave. The SAM density of such a wave can be computed using Equation (1) as:

$$\vec{\mathbf{S}} = \frac{\sigma E_0^2}{2\omega_0} \frac{\vec{\mathbf{k}}}{k}. \tag{3}$$

Thus, the SAM density of a circularly polarized plane wave is uniformly distributed and parallel to the propagation direction of the light field. When such a wave illuminates a light-absorbing or birefringent particle, the SAM is transferred to the particle, inducing rotation along the wavevector $\vec{\mathbf{k}}$. As illustrated in Fig. 1 (b) and (c), a circularly polarized monochromatic plane wave is focused by a lens. Near the focal point, the focused field preserves circular polarization, and its SAM remains aligned with the optical axis. However, when the particle is observed with a camera aligned along the optical axis, only a single projection image from one viewing angle is obtained, leading to the loss of structural information about the particle's three-dimensional geometry, as shown in Fig. 1 (d) and (e). In contrast to directly incident circularly polarized plane wave, we employ a linearly polarized plane wave modulated by a polarization-sensitive spatial light modulator (SLM).

We use a SLM to control a specific directional component of the linearly polarized light field. After the linearly polarized light with zero SAM distribution illuminates the SLM, the outgoing field will have a non-zero SAM distribution. Then a high-numerical-aperture (NA) lens is used to tightly focus the optical field, generating an SAM distribution with an arbitrary direction. Specifically, when the polarization direction of the incident light is oriented at 45° with respect to the polarization axis modulated by the SLM, the phase pattern displayed on the SLM can be controlled according to the following equation:

$$P(\phi) = \text{Arg}(e^{i(\phi - \Phi)} + \frac{a}{2b}\cot\Theta) + \frac{\pi}{2}. \tag{4}$$

Where $\phi$ is the azimuthal angle in polar coordinates, $\Theta$ and $\Phi$ are parameters that describing the direction of the SAM, $a$ and $b$ are parameters related to the NA, and $Arg(\bullet)$ denotes the operator used to calculate the phase of the complex field. Then the SAM of the focused field can be expressed:

$$\bar{\mathbf{S}} = S \begin{bmatrix} \sin\Theta\cos\Phi\bar{\mathbf{e}}_x \\ \sin\Theta\sin\Phi\bar{\mathbf{e}}_y \\ \cos\Theta\bar{\mathbf{e}}_z \end{bmatrix}. \tag{5}$$

Where $\vec{\mathbf{e}}_z$ is the propagation direction of the optical field, $\vec{\mathbf{e}}_x$ is the polarization direction controlled by the SLM and $\vec{\mathbf{e}}_y = \vec{\mathbf{e}}_z \times \vec{\mathbf{e}}_x$. By adjusting the direction of the SAM vector, the particle can be made to rotate about arbitrary axes, thereby enabling the acquisition of multi-view images via a camera that remains coaxially aligned with the optical axis. As shown in Fig. 1 (h) and (i), the transverse SAM induces particle rotation about a transverse axis. and a coaxially aligned camera captures multi-view images. Subsequently, the multi-view images obtained are processed to generate a 3D model of the particle.

**Experiment setup**

To demonstrate the proposed method, we constructed a holographic optical tweezer system based on an inverted microscope (IX73, Olympus), as sketched in Fig. 2. Thus, 3D trapping and rotation of cells can be implemented using a standard setup of holographic optical tweezers. To generate a tightly focused light field with a tunable transverse SAM vector, a linearly polarized beam oriented at 45° is used to illuminate the SLM, which modulates the phase of the incident light field according to Equation (2), which is:

$$P(\phi) = \text{Arg}\left(e^{i(\phi-\Phi)} + \frac{a}{2b}\right) + \frac{\pi}{2}. \tag{6}$$

Then, the trapped cell rotated along the direction of $(90°, \Phi)$ in spherical coordinates, and a CCD camera that was coaxially aligned with the optical axis captured multi-view images of the cell. The obtained images were processed and used to reconstruct the cell's 3D model. In our experiment, we reconstructed 3D models of Punica granatum cell and Prunus cerasifera cell.

**Algorithm workflow based on neural network**

Reconstructing the 3D model of cells from 2D multi-view images typically requires laborious pre-processing and is highly susceptible to degradation caused by defocusing and other environmental disturbances. In order to overcome these problems, we employed a neural network-based pre-processing strategy that exhibits improved robustness and performance, effectively mitigating the impact of slight defocus and environmental noise. The initial stage of our algorithmic pipeline utilizes the YOLOv8 neural network[17], a powerful object detection framework. Using LabelImg, we manually annotated the training and testing datasets to train the model. The trained model was subsequently applied to detect cells in over one thousand images acquired from multiple viewing angles, accurately and effectively extracting regions of interest from the full-field images and cropping them into smaller, consistently sized sub-images. This preprocessing step effectively reduces environmental influence and stabilizes the cell images, thereby facilitating the downstream processing stages. Afterward, traditional binarization techniques, such as Otsu's thresholding, adaptive thresholding, and simple global thresholding, can be applied to extract binary masks from cropped images. However, defocused cell images exhibit diffraction rings, which traditional thresholding and edge detection algorithm often fail to distinguish from the target cells. This limitation leads to poor segmentation performance and significantly degrades the reconstruction quality. To address this issue, we employed the Yolov11-based image segmentation method, which is capable generating accurate binary masks even in the presence of slight defocus. We manually constructed the dataset using LabelMe, which includes both in-focus sharp cell images and slightly defocused images, enabling the network model to better handle defocus artifacts. The trained model demonstrates enhanced robustness, allowing for more accurate segmentation of cells from the background. For the obtained cell segmentation images, we applied rotational alignment to orient the rotational axis along the horizontal or vertical direction, thereby facilitating the subsequent 3D reconstruction process. Following rotation and binarization, we employed the bounding rectangle method to further correct the lateral displacement and stabilize the cell position at the image center. To reconstruct the 3D cell model from the masks obtained at different viewing angles, we employed the visual hull 3D reconstruction algorithm. Each binary defines a 3D volume in space, and the intersection of these volumes from multiple angles yields a 3D representation of the cell. To validate the effectiveness of our method, we performed 3D reconstructions of Prunus cerasifera cells and Punica granatum pollen cells. Fig. 4 (a) shows the reconstructed 3D model of the Punica granatum pollen cell, and Fig. 4 (b) corresponds the Prunus cerasifera cell. To better visualize the reconstructed 3D model and reveal structural features, we used color mapping to represent the distance from each point on the cell surface to the centroid of the model, as shown in Fig. 4(a) and Fig. 4(b). As shown in Fig. 4 (a), the reconstructed Punica granatum pollen cell exhibits three prominent protrusions evenly distributed on the surface. The color gradient indicates that the most significant depth variations occur at these three regions, while the rest of the surface remains relatively uniform, with predominantly blue tones representing minimal deviation from the centroid. Aside from these distinct apex features, the overall morphology of the Punica granatum pollen cell appears relatively spherical and symmetric. In contrast, the Prunus cerasifera cell exhibits a more irregular morphology, characterized by an overall ellipsoidal shape with distinct proximal and distal ends, as shown in Fig. 4 (b). Furthermore, we present the original multi-view images of both cells at various rotation angles, their corresponding segmentation masks, and the projected images of the reconstructed 3D models. The masks and model projections show strong agreement with the raw images. Additionally, Fig. 4 (c) and Fig. 4 (d) display the multi-view and cross-sectional visualizations of the two cells. In our setup, one pixel in the obtained images corresponds to 0.1372 μm. Consequently, each voxel in the reconstructed 3D model corresponds to a volume of 0.1372 ×0.1372×0.1372 μm$^3$, which defines the isotropic spatial resolution of the system. Then, the cell volume was directly calculated by summing the total number of occupied voxels and multiplying by the volume of a single voxel. The calculated volume of the Punica granatum pollen cell is 6021.3 μm$^3$, while the volume of the Prunus cerasifera is 1173.1 μm$^3$.

**DISCUSSION**

In summary, we have used single-beam holographic optical tweezers to achieve stable three-dimensional rotation of cells, and employed a coaxially aligned camera to acquire multi-view images, enabling in situ isotropic 3D reconstruction. As a demonstration, we employed tightly focused light field with transverse SAM to trap and rotate Prunus cerasifera and Punica granatum pollen cells, while multi-view images were acquired using a camera aligned with the optical axis. Through a neural network-based workflow, we reconstructed the 3D model of the cells from multi-view images with isotropic resolution. In our approach, a single focused beam simultaneously provides both the restoring force and the torque for trapping and rotation.

Conventional 3D imaging methods are often constrained by limited viewing angles, which can lead to the loss of 3D structural information of the cells and a decrease in axial resolution, a phenomenon known as the "missing cone". In contrast, our system provides a simple and effective method to acquire full-angular information of cells, thereby enabling isotropic resolution. Compared to traditional methods, our manipulation approach is more flexible and stable, while also requiring a simpler experimental setup. By integrating a neural network-based workflow, we achieved more accurate cell image segmentation, which effectively reduces background interference and performance reliably even on slightly defocused images. Furthermore, thanks to the simplicity of the setup and the flexibility and stability of the manipulation, our method can be applied to a broader range of fields.

**METHODS**

**Details of the experimental setup.** As shown in Fig. 2, our setup is a standard holographic optical tweezers system based on an inverted microscope (IX73, Olympus). A linearly polarized laser beam is generated from 532 nm continuous-wave solid-state laser (MSL-R-532-5W, Changchun New Industries, Ltd.). Then, a 4f-configuration consisting of L1 and L2 is used to expand the laser beam, ensuring complete illumination of the liquid crystal SLM (1920×1080-pixel resolution; Pluto, HOLOEYE). A half-wave plate is placed in front of the SLM to tune the polarization of the incident light, ensuring that the polarization of the light is at a 45° to the polarization direction controlled by the SLM. Thus, only the x-component of the incident field, which is controlled by the SLM, will be manipulated, while the y-component remains unchanged. Through modulate the phase distribution on the SLM, the outgoing light field will have different SAM distributions. Then the modulated light field is relayed to the pupil plane of an oil-immersion objective lens (100×, NA=1.3; UPLFLN100XO2, Olympus). Finally, the optical trap generated by the focused beam can both confine and rotate cells within a sealed sample cell. Additionally, the movements of the trapped cells are monitored in real time using the microscope and a CMOS camera (PL-D752MU, PixelLINK).

# REFERENCES

[1]     J. Lim, K. Lee, K.H. Jin, S. Shin, S. Lee, Y. Park, J.C. Ye, Comparative study of iterative reconstruction algorithms for missing cone problems in optical diffraction tomography, Opt Express, 23 (2015) 16933-16948.

[2]     B. Simon, M. Debailleul, M. Houkal, C. Ecoffet, J. Bailleul, J. Lambert, A. Spangenberg, H. Liu, O. Soppera, O. Haeberlé, Tomographic diffractive microscopy with isotropic resolution, Optica, 4 (2017).

[3]     J. van Rooij, J. Kalkman, Large-scale high-sensitivity optical diffraction tomography of zebrafish, Biomed Opt Express, 10 (2019) 1782-1793.

[4]     K.C. Zhou, R. Qian, S. Degan, S. Farsiu, J.A. Izatt, Optical coherence refraction tomography, Nat Photonics, 13 (2019) 794-802.

[5]     K. Kim, J. Yoon, Y. Park, Large-scale optical diffraction tomography for inspection of optical plastic lenses, Optics Letters, 41 (2016) 934-937.

[6]     M. Kvale Lovmo, S. Deng, S. Moser, R. Leitgeb, W. Drexler, M. Ritsch-Marte, Ultrasound-induced reorientation for multi-angle optical coherence tomography, Nat Commun, 15 (2024) 2391.

[7]     D. Ahmed, A. Ozcelik, N. Bojanala, N. Nama, A. Upadhyay, Y. Chen, W. Hanna-Rose, T.J. Huang, Rotational manipulation of single cells and organisms using acoustic waves, Nat Commun, 7 (2016) 11085.

[8]     J.H. Lee, J.W. Kim, M. Levy, A. Kao, S.H. Noh, D. Bozovic, J. Cheon, Magnetic Nanoparticles for Ultrafast Mechanical Control of Inner Ear Hair Cells, Acs Nano, 8 (2014) 6590-6598.

[9]     Y. Mao, S. Li, Z. Wang, M. Shao, P. Wang, X. Tan, F. Lu, Y. Wang, X. Wei, Z. Zhong, J. Zhou, Optofluidic-based cell multi-axis controllable rotation and 3D surface imaging, Applied Physics Letters, 123 (2023).

[10]    J. Sun, B. Yang, N. Koukourakis, J. Guck, J.W. Czarske, AI-driven projection tomography with multicore fibre-optic cell rotation, Nat Commun, 15 (2024) 147.

[11]    M. Lee, K. Kim, J. Oh, Y. Park, Isotropically resolved label-free tomographic imaging based on tomographic moulds for optical trapping, Light: Science & Applications, 10 (2021) 102.

[12]    F. Merola, L. Miccio, P. Memmolo, G. Di Caprio, A. Galli, R. Puglisi, D. Balduzzi, G. Coppola, P. Netti, P. Ferraro, Digital holography as a method for 3D imaging and estimating the biovolume of motile cells, Lab Chip, 13 (2013) 4512-4516.

[13]    L. Peng, L. Duan, K. Wang, F. Gao, L. Zhang, G. Wang, Y. Yang, H. Chen, S. Zhang, Transverse photon spin of bulk electromagnetic waves in bianisotropic media, Nature Photonics, 13 (2019) 878-882.

[14]    A. Aiello, P. Banzer, M. Neugebauer, G. Leuchs, From transverse angular momentum to photonic wheels, Nature Photonics, 9 (2015) 789-795.

[15]    Y.J. Wu, J.H. Zhuang, P.P. Yu, Y.F. Liu, Z.Q. Wang, Y.M. Li, C.W. Qiu, L. Gong, Time-varying 3D optical torque via a single beam, Nat Commun, 16 (2025) 593.

[16]    A. Laurentini, The Visual Hull Concept for Silhouette-Based Image Understanding, Ieee T Pattern Anal, 16 (1994) 150-162.

[17]    D. Reis, J. Kupec, J. Hong, A. Daoudi, Real-time flying object detection with YOLOv8, arXiv preprint arXiv:2305.09972, (2023).

**Authors Contributions**

L.G. conceived the project. Y.W. designed the system, and Y.P. performed experiments and analyzed the experimental results. Y.P., and Y.W. wrote the manuscript. All authors were involved in revising the manuscript.

**Additional Information**

Supplementary information is available in the online version of the paper. Correspondence and requests for materials should be addressed to L.G.

**Competing financial interests:** The authors declare no competing financial interests.

**Data and materials availability:** Data supporting the findings in this study are available within the article and its supplementary information files and from the corresponding author upon reasonable request.

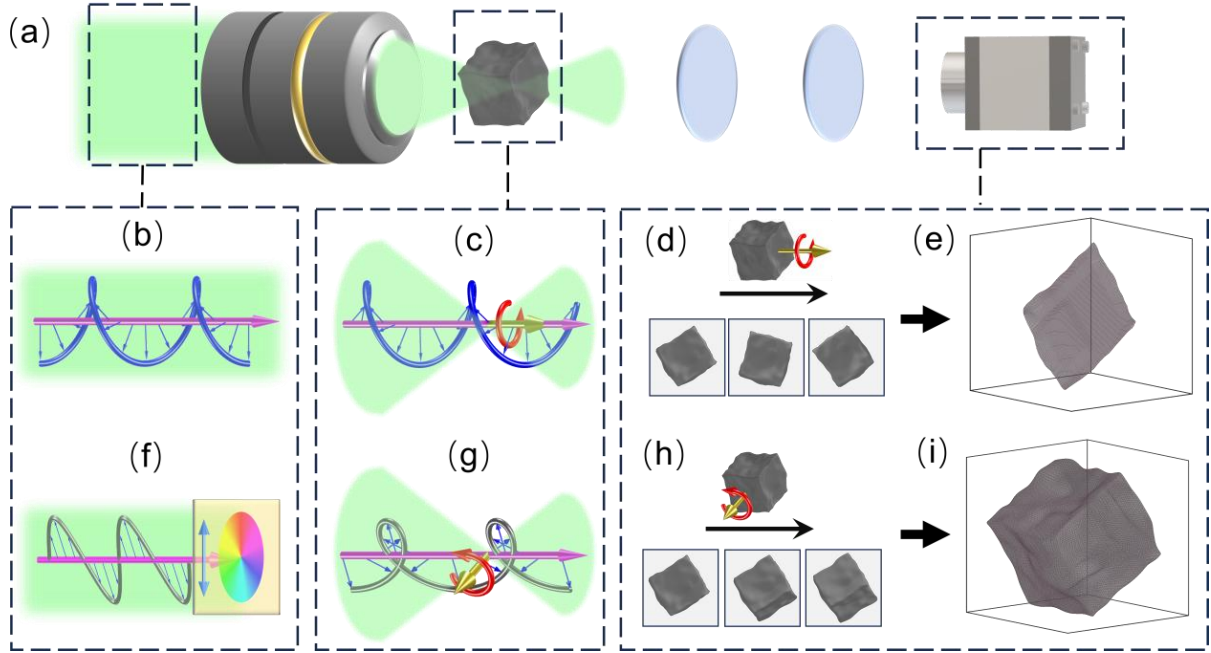

**Fig. 1 Concept and principle of our system.** (**a**) The conceptual diagram of the setup shows that a high-NA objective tightly focuses the light field to trap and rotate the particle, while a coaxially aligned camera captures its rotation. (**b**) The incident beam is a circularly polarized plane wave. (**c**) After tight focusing, the circularly polarized light maintains its polarization near the focal point, with the SAM direction aligned parallel to the optical axis. (**d, e**) With longitudinal SAM, the particle rotates along the optical axis, and the camera captures only a single-angle projection, leading to the loss of multi-angle spatial information. (**f**) The incident beam is linearly polarized, with its polarization direction oriented at a 45° relative to SLM-controlled direction. (**g**) By modulating the phase distribution on the SLM, the tightly focused field has transverse SAM. (**h, i**) Spatial information of the particles is obtained from multi-view images, enabling the reconstruction of their 3D models.

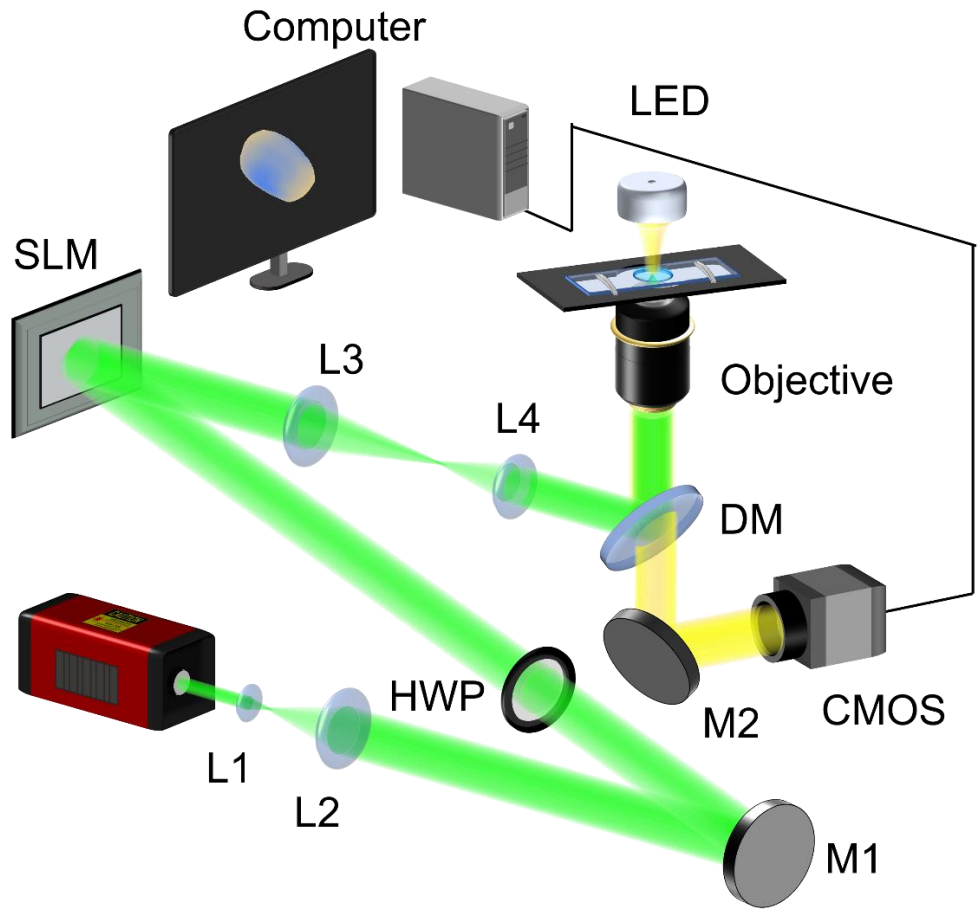

**Fig. 2 Experimental setup** L1-L4: lens; M1, M2: mirror; HWP: half-wave plate; SLM: spatial light modulator; DM: dichroic mirror.

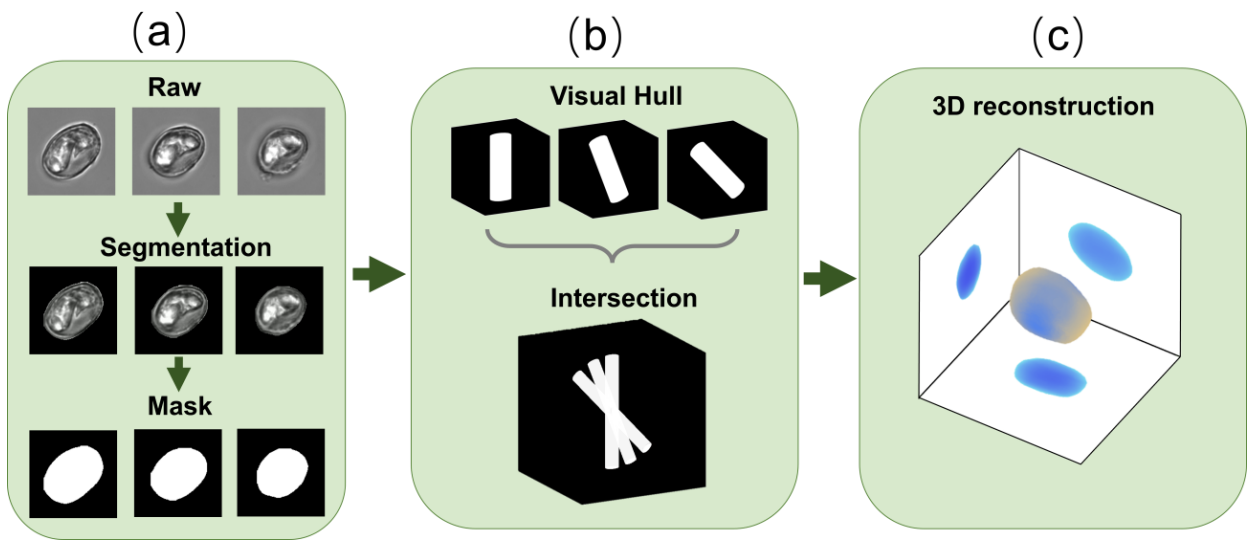

**Fig. 3 3D reconstruction workflow.** (**a**) A neural network-based method is employed to perform cell segmentation, enabling the extraction of multi-angle cell images from the background while simultaneously obtaining the corresponding masks. (**b**)The visual hull method is used for 3D reconstruction. Each cell mask obtained from different viewing angle defines a possible 3D spatial region, and the intersection of these regions forms the 3D cell model. (**c**) The reconstructed 3D cell model by our proposed method.

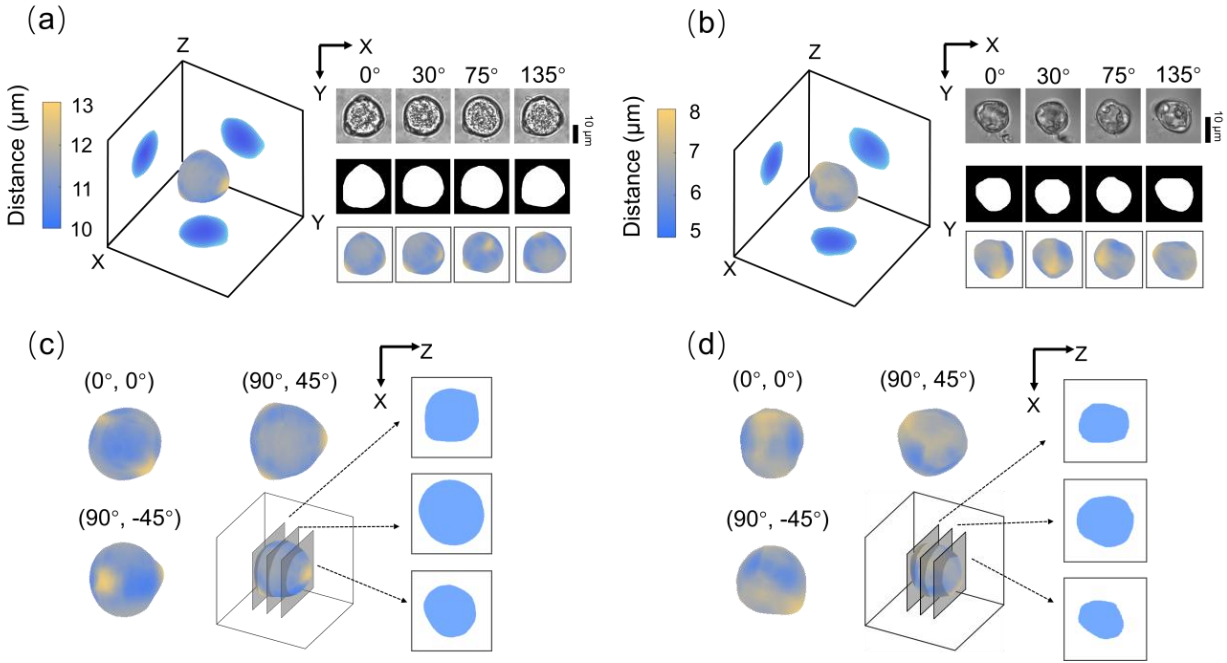

**Fig. 4 Experimental demonstration of our method.** (**a, b**) The reconstructed 3D model of the Punica granatum pollen cell and Prunus cerasifera cell are shown alongside the raw multi-angle images, the corresponding binary mask images, and the projections of the reconstructed 3D model. The color represents the distance from each surface point to the cell's centroid. (**c, d**) Three-view projections of the reconstructed model, accompanied by cross-sectional slices taken at different depths.